# Individual Pulse Monitoring and Dose Control System for Pre-Clinical Implementation of FLASH-RT

M. Ramish Ashraf[1], Mahbubur Rahman[1], Xu Cao[1], Kayla Duval[2], Benjamin B. Williams[1,2,3], P. Jack Hoopes[1,3,4], David J. Gladstone[1,2,3], Brian W. Pogue[1,3,4], Rongxiao Zhang[1,2,3], Petr Bruza[1]

[1] Thayer School of Engineering, Dartmouth College, Hanover NH 03755, US

[2] Department of Medicine, Geisel School of Medicine, Dartmouth College Hanover NH 03755 USA

[3] Norris Cotton Cancer Center, Dartmouth-Hitchcock Medical Center, Lebanon, NH 03756 USA

[4] Department of Surgery, Geisel School of Medicine, Dartmouth College, Hanover NH 03755 USA

*Corresponding Authors: Petr Bruza (petr.bruza@dartmouth.edu) and M. Ramish Ashraf (ramish.th@dartmouth.edu)

**Abstract:**

Ultra-high dose rate electron sources require dose rate independent dosimeters and a calibrated dose control system for accurate delivery. In this study, we developed a beam monitoring and control system based on fast processing and characterized a dose-rate independent scintillating detector at ultra-high dose rates. A commercially available point scintillator detector was coupled to a gated integrating amplifier and a real-time controller for dose monitoring and feedback control loop. The controller was programmed to integrate dose and measure pulse width of each radiation pulse and provide feedback to the linac when the prescribed number of pulses or dose was delivered. The Scintillator was mounted in solid water phantom for dose feedback, and placed underneath mice skin for in vivo monitoring. Additionally, the scintillator was characterized in terms of its radiation stability, mean dose-rate ($\dot{D}_m$), and dose per pulse ($D_p$) dependence. Dose integration was performed for each radiation pulse and displayed in real-time. The $D_p$ and the pulse width showed a consistent ramp-up period across ~4-5 pulse. The scintillator was shown to be linear with $\dot{D}_m$ (40-380 Gy/s) and $D_p$ (0.3-1.3 Gy/Pulse) to within +/- 3%. However, the plastic scintillator was subject to significant radiation damage (16%/kGy) for the initial 1kGy and would need to be calibrated frequently. Pulse-counting control was accurately implemented with one-to-one correspondence between the intended and the actual delivered pulses. The dose-based control was sufficient to gate on any pulse of the linac. In-vivo dosimetry monitoring with a 1 cm circular cut-out revealed that during the ramp-up period, the average Dp was ~0.045 ± 0.004 Gy/Pulse, whereas after the ramp-up it stabilized at 0.65 ±0.01 Gy/Pulse. The tools presented in this study can be used to determine the beam parameter space pertinent to the FLASH effect. Additionally, this study is the first instance of real-time dose-based control for a modified linac at ultra-high dose rates.

## 1. Introduction:

FLASH is a novel ultra-high dose-rate radiation therapy technique, which has gained significant interest due to its ability to spare some normal tissues as compared to conventional forms of radiotherapy[1] while still providing effective tumor control. The FLASH effect has now been observed in multiple animal and organ[2–4] models. Moreover, the first human patient was treated using FLASH[5] and a clinical human trial is currently in progress[6]. One of the factors hindering pre-clinical and eventual clinical adaption of FLASH is the issue of performing accurate dosimetry and real-time beam monitoring at ultra-high dose-rate conditions[7–12]. It is critical that the pre-clinical studies being performed have accurate dosimetry, because the radiobiological underpinnings of FLASH are not currently well understood and accurate dosimetry will be needed to build this understanding.

There is an emergence of modified clinical accelerators which enable delivery of beam in the UHDR regime (i.e. > 40 Gy/s)[13–15]. At our institution, a clinical LINAC was modified by disabling feedback mechanisms and retracting the target to deliver a 10 MeV UHDR beam, with a dose per pulse of 1 Gy at iso-center[15]. Conventional linear accelerators rely on ionization chambers for the purpose of real-time beam monitoring and feedback based on beam flatness, symmetry, dose rate and total delivered dose. However, ionization chambers are known to be dependent on dose-rate due to ion-recombination issues at high instantaneous dose-rates, although recombination models have suggested that are valid up to 2-3 Gy/Pulse.[16–18] Due to this dose rate dependence, most of the aforementioned feedback systems were disabled to realize the FLASH beam. Even if the internal LINAC feedback systems were to be used, they lack the ability to offer single pulse temporal resolution, which is critical for accurate FLASH studies. In most cases, external pulse counting circuitry has been employed as a beam control system. The current beam control system at our institution relies on a remote trigger unit (RTU) interfaced with an Arduino Controller, following the work of Schuler et al[14]. The RTU consists of two scintillators coupled with silicon photomultiplier (SiPM) detectors which are used to detect stray radiation inside the room and produce a TTL output only where there is a radiation beam present[19]. The two-scintillator configuration is used for coincidence detection to minimize detection of spurious pulses due to neutron activation and or cosmic rays. Spurious pulses here refer to pulses that are not representative of the pulse produced from the linac but are still registered as valid pulses by the pulse counting circuity. Finally, the output from the Arduino controller is fed to Varian's Beam Gating switchbox (Varian, Palo Alto, CA). With external pulse counting circuitry, there is significant uncertainty in dose delivery because a ramp-up period is observed where the first 4-5 pulses underdose significantly. The ramp-up period can itself be variable in terms of how many pulses are required to achieve beam stability. With a dose per pulse near 1 Gy, any variability in the ramp-up period results in significant uncertainty in the dose delivery for low doses. Additionally, it has also been observed that the number of pulses produced by the RTU depends highly on the positioning of the detector and the type of beam collimation being used. If the detector is placed far away from the projected field size, one can potentially run into the issue of missing pulses due to insufficient stray radiation. An obvious solution would be to place the RTU closer to the beam; however, if the detector is placed too close to the projected field size, spurious pulses, not representative of the actual radiation pulses produced by the linac, can be observed due to scintillator afterglow, and/or the neutron activated products that decay at their own rate. Of note, it has been observed that the issue of neutron activation is prevalent with electron applicators and is less of a problem with an uncollimated beam. Additionally, extended beam-on periods also increase the probability of spurious pulses. Although, coincidence circuitry is implemented in the trigger unit, there are still occasional spurious pulses that end up being counted as actual radiation pulses. These issues contribute to the uncertainty in the dose delivery and point towards the need for hardware instrumentation and dose-rate independent detectors that can measure dose on a pulse-by-pulse basis and use the accumulated dose as a feedback signal to the linac.

Motivated by mitigating uncertainty in dose delivery in UHDR operation, this study introduces and characterizes detectors and hardware for online monitoring and the first-ever dose-based control for a FLASH beam. Two redundant forms of feedback (pulse-counting and dose-accumulation) to the linac were realized and characterized. The scintillator system is compared against GafChromic film; a detector which has been established to be dose-rate independent by multiple investigators[8,20]. The study is also motivated by the need to have techniques that can accurately probe the temporal structure of the beam. It is hypothesized that quantities such as the repetition rate and dose per pulse play a crucial role in successful elicitation of the FLASH effect and therefore by measuring these quantities for each beam delivery, the FLASH effect can be understood in terms of its relation to the temporal structure of the beam.

For the purposes of real-time dose accumulation and feedback to the linac, a field programmable gate array (FPGA) based embedded industrial controller was used. FPGA circuits typically consist of discrete logic units such as lookup tables and logic gates, which are connected via programmable interconnects. These FPGAs are advantageous for highly deterministic tasks because each independent processing task can be assigned to a dedicated section on the chip and can function independently of other tasks with minimal jitter. As mentioned earlier, two independent feedback tasks were accomplished via the embedded controller: 1) pulse counting, and 2) dose-based feedback. For probing dose, a commercially available point scintillating detector (PSD), Exradin W1 (Standard Imaging, Middleton, WI) was employed. The W1 PSD has a diameter of 1.0 mm and is 3 mm long with a physical density of 1.05 g/cm$^3$. The detector itself is packaged in a housing which has a 2.8 mm diameter and is 42 mm long. Scintillators are a viable choice for FLASH due to their dose-rate linearity, high spatio-temporal resolution (~ ns and ~mm) and water equivalence[21]. One drawback of scintillator detectors is the production of Cherenkov radiation and/or fluorescence in the light guide. There have been numerous publications highlighting different Cherenkov discrimination techniques[22,23]. However, if calibrated properly, the W1 can be accurately used for clinical dosimetry and can even serve as a reference detector for small field correction factors for other non-ideal detectors[24].

## 2. Methods and Materials:

### 2.1. Real-Time Beam Monitoring Hardware

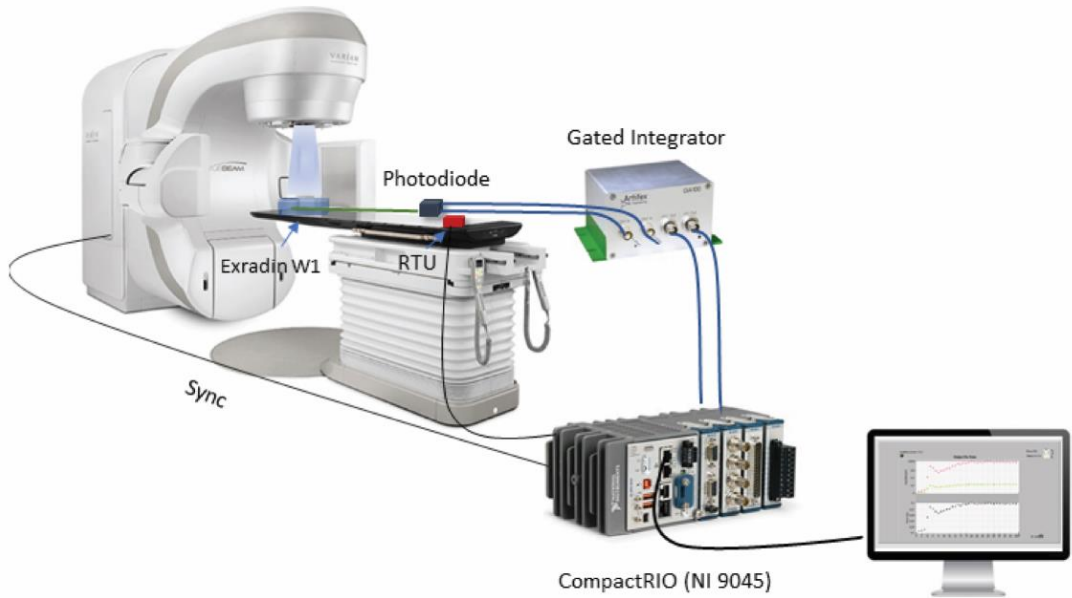

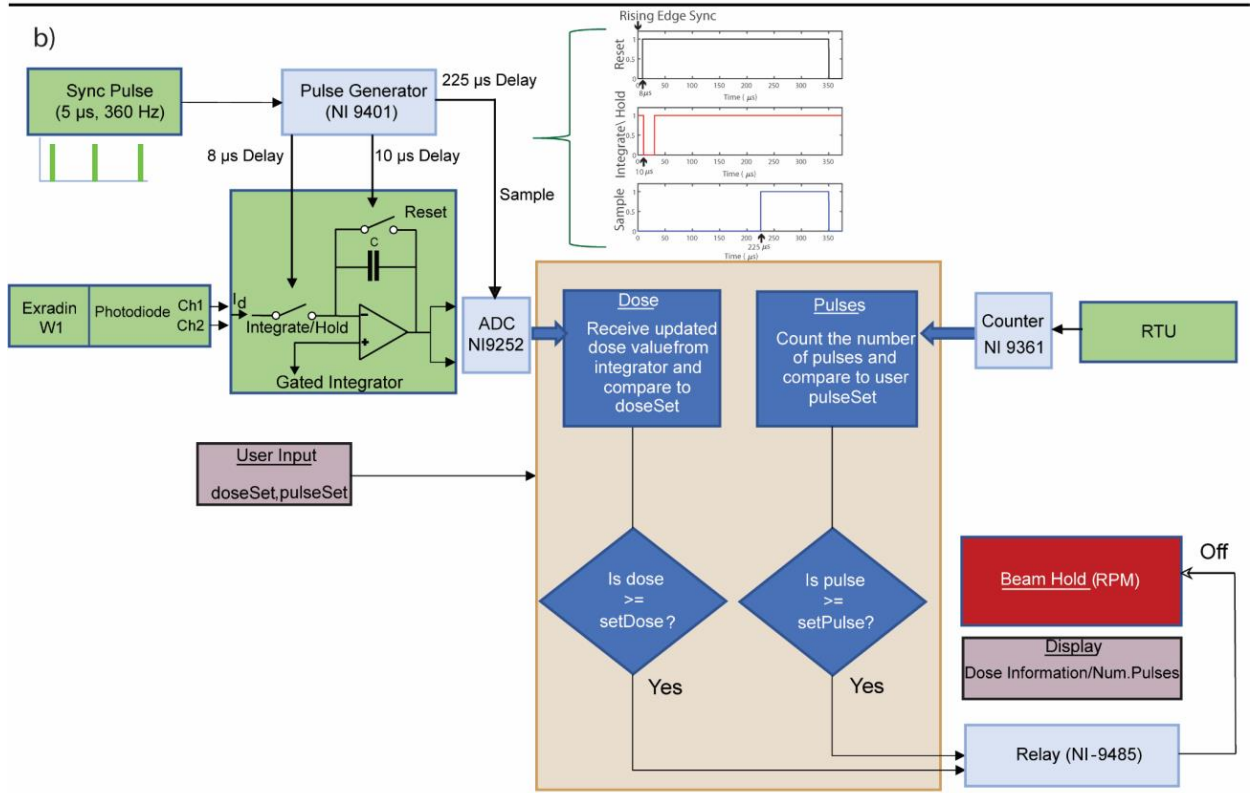

**Figure 1a)** Schematic of the hardware and signal routing for real-time beam monitoring. The beam monitoring hardware primarily consisted of a real-time controller, a gated integrator, a plastic scintillator, and a remote trigger unit. **b)** Shows the control algorithm consisting of redundant dose feedback based on pulse counting and dose accumulation. The signal routing is also depicted here along with the different I/O modules and radiation detectors. The pulse sequence realized via the hardware to integrate dose per pulse is also shown in b). Green colored boxes represent external signals from the linac and detectors, whereas the light blue boxes represent the I/O modules used in this study. The yellow box in the middle denotes the control algorithm used in this study. Boxes with light shade of red denote the information required from the user and displayed after dose delivery.

Figure 1 depicts the hardware, detectors, and signal routing for pulse-resolved beam monitoring. There are three major components of the hardware solution: 1) A Compact Real-Time Input/Output (CompactRIO NI 9045) controller, 2) External Detectors, and 3) Trigger signals to and from the LINAC. These components and their interaction are explained next. A real-time FPGA based controller, (CompactRIO, NI-9045) manufactured by National Instruments (Austin, TX) was used for online beam monitoring. The controller consisted of an Kintex-7 70T FPGA board which interfaces directly with input/output (I/O) modules connected to external detectors. The FPGA was programmed through the LabView graphical programming language using the LabView FPGA module. The controller chassis provided 8 slots for interfacing with different I/O modules, making it a highly modular and customizable system. Additionally, included with the controller is a dedicated real-time operating system (RTOS) which was useful for floating point operations and high throughput streaming of data, i.e., processes which would otherwise take up significant resources on the FPGA circuit. The RTOS was also programmed via the LabView environment. The RTOS displayed real-time data communicated from the FPGA, mainly information about the dose per pulse and the pulse width. The FPGA portion of the hardware was programmed to accomplish three major tasks which were implemented physically on different parts of the FPGA chip, thus ensuring true parallel control.

**2.1.1 Synchronization & signal sampling.** The first task implemented on the hardware was to provide signals to the gated integrator which initiated with input from the 'Sync' signal from the linac and generated signals for a gated integrator, which is described briefly. The 'Sync' signal consisted of 5 µs bursts with a repetition rate of 360 Hz and was typically used as a trigger source for automatic frequency control (AFC) and oscilloscope monitoring of multiple LINAC waveforms. The radiation pulse occurred 12 µs after the rising edge of the Sync pulse. Taking these timing considerations into account, the FPGA was programmed to generate reset, integrate/hold, and sample signals for the gated integrator with delays of 8 µs, 10 µs and 225 µs, respectively from the leading edge of the Sync signal, as shown in Figure 1. These timing signals resulted in a dose integration period of ~ 20 µs. The time delays were optimized by generating artificial signals mimicking the linac pulse structure from a function generator and taking into consideration the settling time of the integrator.

**2.1.2 Dose integration**. The second task was to perform dose integration per pulse and sample the dose value as triggered by a signal delayed by 225 us from the rising edge of the sync pulse and sends a beam gating signal to the Linac via the relay if the desired dose value is reached. The integration of dose within each pulse is realized via a dual channel gated integrator, GIA100 (Artifex, Emden, Germany). The gated integrator acts as simple integrator (i.e., a capacitor in the feedback loop of an operational amplifier) with the additional ability to gate the integrating process if provided with the appropriate timing signals. A schematic of a generic gated integrator is shown in Figure 1b), where it is connected to the output from the W1's photodiode. If the reset switch was open and the integrate/hold switch was closed, a current source would charge the capacitor C. Since the inverting input acts as a virtual ground, the output is given by:

$$V_o = \frac{1}{C} \int I_d(t) dt \qquad (1)$$

After integration, the voltage is held constant for a specified duration and sampled by the ADC until the reset switch is closed, after which the capacitor discharges and is ready to integrate the next pulse. The GIA100 employs a 100 pF capacitor and the nominal switching time is ~ 1 µs. The gated integrator essentially replaces the role of the electrometer that would otherwise be typically used with point detectors for signal accumulation in conventional radiotherapy dosimetry. A dual channel gated integrator is required to acquire data in different wavelength bands for removal of the stem effect in the Exradin

W1. The Exradin W1 is equipped with a beam splitter and two photodiodes, which acquire signals in two different wavelengths bands that can be used to suppress the Cherenkov stem effect using a chromatic removal technique[25,26].

**2.1.3 LINAC control.** Finally, the third task was to provide control to the linac primarily in terms of dose accumulation, but pulse counting was also used as a secondary form of control. The dose accumulation process is described above, and pulse counting was achieved using RTU as the external detector. The code accounts for spurious pulses by ensuring that only those pulses are counted which produce a TTL output which is high for at least 1 µs. The pulse counting and dose counting routines send a beam gating signal to the linac via Varian's Real-time Position Management (RPM) gating box, if the required number of pulses or dose, respectively have been reached. but is primarily meant to be used as secondary form of feedback, with the primary feedback dependent on dose accumulation.

**2.1.4 Pulse Counting and Pulse Width.** Finally, the fourth task was to detect the radiation pulses from the RTU and measure pulse width of each pulse and report relevant temporal metric to the user.

The different I/O modules used in this study to accomplish the aforementioned tasks are briefly described in the supplementary section.

**2.2 Dose Calibration**

As mentioned earlier, Cherenkov radiation and/or fluorescence produced in the light guide can be a source of large uncertainty for any fiber optic-based dosimeter. The stem signal is typically proportional to the amount of light guide/fiber optic in the radiation beam. Therefore, stem signal is only a cause of concern if the irradiation conditions change in terms of how much of light guide is irradiated. For beam monitoring purposes, one would ideally want to keep the fiber optic at a constant position with respect to the beam; in such a case removal of the stem signal might not be required. However, even with a constant fiber position, beam steering issues and different beam collimation devices would result in different stem contributions to the signal. Therefore, for robust dose calibration and the potential to perform in-vivo dosimetry, the stem removal was included. The stem removal in the Exradin W1 is implemented according to technique presented by Fontbonne et al[25] and Frelin et al[26] and entails measurement of two factors; 1) gain and 2) Cherenkov light ratio (CLR). To determine the calibration coefficients 5 repeated measurements were performed with the W1 placed at iso-center (100 cm SAD) for both the minimum and maximum fiber configuration. A total of 35 linac pulses were delivered and a 40 x 40 cm$^2$ jaw size was used. Additionally, a 1 cm build up solid water slab was placed on top of W1. In each case, the total dose delivered was verified using GafChromic film. FilmQA$^{TM}$ (Ashland, Wilmington, DE) software, which employs the triple channel dosimetry formalism with non-uniformity correction[27], for calibrating and reading-out the films.

**2.3 Exradin W1 Characterization**

Once the calibration coefficients were obtained, a series of experiments were performed to characterize the W1 under ultra-high dose-rate conditions. For each study, 35 linac pulses were delivered, and the measurements were repeated three times, unless otherwise stated. Dose verification was performed for each beam delivery using GafChromic film.

### 2.3.1) Dose-Rate Dependence

The dose-rate dependence of the detector was evaluated by either changing the repetition rate of the linac or changing the source to surface distance (SSD). The former case tests the mean dose-rate dependence, and the latter case tests the instantaneous dose-rate dependence (i.e., dose-rate within the pulse) of the detector. For Varian based linacs, the radiation beam is typically delivered in 5 µs bursts at a repetition rate of 360 Hz (600 MU/min). To achieve lower average dose-rates, pulses are dropped by delaying the gun pulse relative to the accelerating klystron pulse. In this study, repetition rates of ~360 Hz, 240 Hz, 120 Hz and 60 Hz were used to test the mean-dose rate dependence. For the instantaneous dose-rate study, four different SSDs were used: 77 cm, 90 cm, 110 cm, and 130 cm. To note, no beam collimation was used for this study.

### 2.3.2) Field Size Dependence

For most pre-clinical FLASH studies, small field sizes are typically used for irradiating mice and other small animals. The Exradin W1 has been shown to be an ideal small field dosimeter, requiring minimal correction factors. Therefore, it is expected that the small field sizes should present no issues with the W1. The purpose of this test was to primarily assess if the Cherenkov discrimination was implemented accurately, since changing field size from the calibration conditions (40 x 40 cm$^2$) to smaller field sizes results in a vastly different amount stem irradiation. Therefore, 4 cm ⌀, 3 cm ⌀, 1.5 cm ⌀ and a 1 cm ⌀ circular beams were delivered to a solid water slab with the W1 at a SAD of 100 cm.

### 2.3.3) Radiation Damage

If the W1 is meant to replace the transmission chamber as a permanent beam monitoring device, it is critical that radiation damage endured by the W1 and its effect on the sensitivity of the detector is quantified. The radiation damage response of the W1 was quantified by keeping track of the sensitivity over an extended period and periodically recalibrating the CLR and gain values to quantify the change in sensitivity.

### 2.3.4) In-Vivo Dosimetry

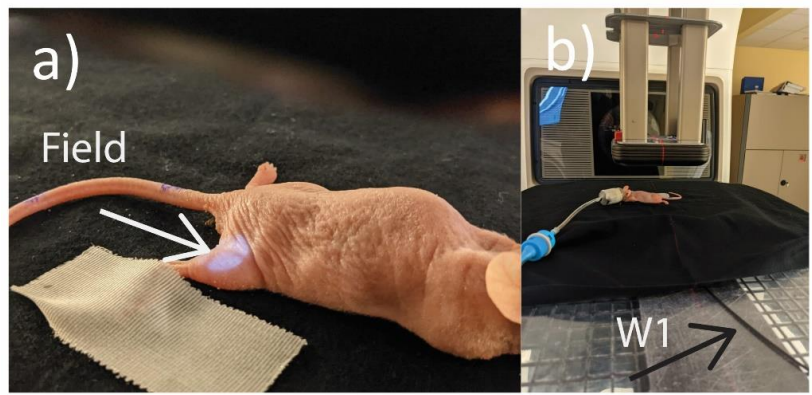

**Figure 2 a) and b)** In-Vivo Dosimetry Setup for the mice experiments, with the mouse lined up in the beam (as shown by the field light) underneath the applicator.

As an example of the new beam monitoring solution, the hardware presented in this study was used for in-vivo pulse-resolved dosimetry of murine experiments and verified against GafChromic film. All animal procedures were approved by the Dartmouth Institutional Animal Care and Use Committee (IACUC) and the work in this paper followed these approved procedures throughout the study. A small 1 cm ⌀ circular field was used to irradiate 4 different mice around the leg region. Mice were placed on top of the calibration solid water slab provided by Standard Imaging (Figure 2). 40 pulses were delivered such that the final cumulative dose was ~ 20 Gy to the surface of the leg. The W1 was used to measure the exit dose. Additionally, film was placed underneath the mice and on top of the W1 to serve as a reference.

## 3. Results:

### 3.1. Pulse Counting Feedback

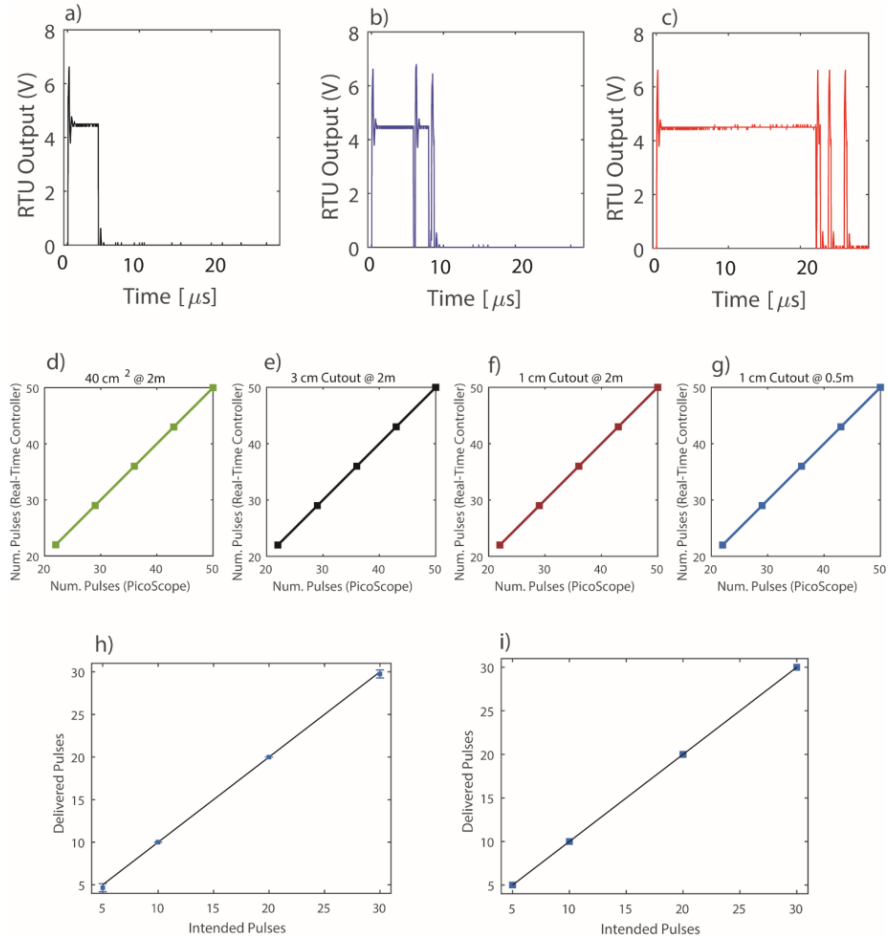

**Figure 3 a, b, and c)** Depicts the signals generated from the RTU with the RTU placed at 2 m for a 40 cm², 1 cm cutout and 0.5 m for the 1 cm cutout, respectively. **d,e,f)** shows the one-to-one correspondence between number of pulses counted by the hardware presented in this study versus an offline comparison with a digital oscilloscope. **h) and i)** show the feedback to the linac in terms of pulse counting.

The pulse counting implementation was tested by placing the RTU at different locations from the iso-center and using different field sizes or beam collimation. The issue of spurious pulses is depicted in Figure 3 a), b) and c). In particular, Figure 3a) shows the RTU output for a completely open field with the RTU at 2 m from the iso-center. However, as beam collimation is increased in Figure 3b) (i.e., 1 cm cutout), the issue of spurious pulses after the main radiation pulse is evident. Moving the RTU closer to the field, exacerbates the issues as seen in Figure 3c). Note that moving the RTU closer to the projected field and/or adding beam collimation also results in a larger pulse width than what is expected of a typical linac pulse. This is because of the abundance of stray radiation closer to the source and the decay kinetics of the BGO (~ 300ns decay time + afterglow) scintillator. The pulse counting routine implemented in the FPGA accounts for these spurious pulses by detecting the rising edge of only those peaks which have a minimum pulse width of 1 µs, thereby rejecting the high frequency noise. The one-to-one correspondence between the number of pulses counted by this technique versus an offline comparison with a digital oscilloscope is shown Figure 3 d, e and f) for different beam collimations and RTU distances. Of note, for this test, the built-in transmission chamber was used as beam-off signal. Increasing number of monitor units were delivered from the linac console. With a repetition rate of 360 Hz (i.e., 600 MU/min), the ratio between MU and the number of pulses was 1:7. Note that, although the transmission chamber suffers from ion-recombination at high dose-rates, for a signal repetition rate the output of the transmission chamber was found to be stable.

The pulse counting feedback circuit was tested by disabling the transmission chamber control and the linac was controlled using the pulse counting routine implemented on the hardware. The RTU was again placed at 0.5 m from iso-center and a circular cutout of 1 cm was used to maximize the issue of spurious pulses. Two different pulse repetition rates were used (360 Hz and 60 Hz) to verify that the pulse counting feedback was valid over the complete range of reputation rates. The results for 360 Hz and 60 Hz are displayed in Figure 3 h) and i), respectively. There was one to one correspondence between the intended number of pulses and actual pulses delivered. The experiment was repeated 5 times and the standard deviations are plotted on the graphs in Figure 3 h and i); for the 60 Hz, the delivered pulses were exactly equal to the intended pulses in each cases (i.e., 0 standard deviation). However, for 360 Hz, one measurement each for the 5 and the 30 pulse delivery, delivered one less pulse (4 and 29). All other measurements had exact one-to-one correspondence.

## 3.2 Dose Integration and Calibration

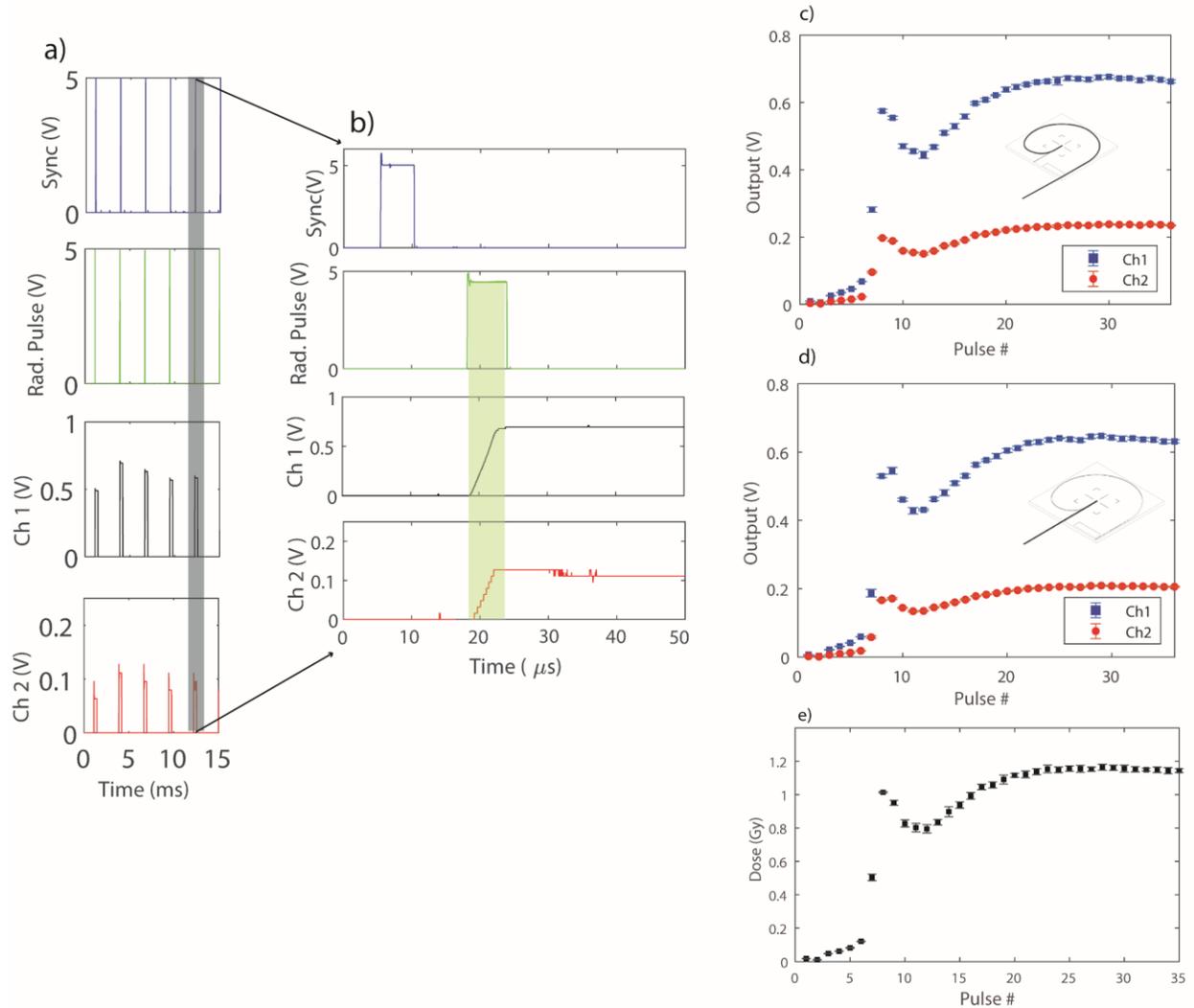

**Figure 4 a)** Shows a waveform capture of the sync signal, radiation pulse and the integrated two channel output of the photodiode for a sample irradiation. **b)** Shows a zoomed-in portion of Figure 5a, depicting the dose integration for an individual linac pulse (shaded in green). **c,d)** Output from the Exradin-W1 photodiode for the maximum and minimum configuration, respectively. **e)** Depicts a sample dose delivery of 35 pulses. The error bars for each graph represent the standard deviation from 5 repeated measurements.

Figure 4 depicts the different waveforms generated and captured during the dose delivery process. In particular, Figure 4 a and b) show the dose accumulation process for a typical FLASH irradiation. It can be seen that dose integration takes place for the two outputs of the photodiode synchronously with the radiation pulse. The integrated dose is then reset before the onset of the next Sync pulse. Channel 1 (Ch 1) here denotes the scintillation signal (green wavelength band), whereas channel 2 (Ch 2) denotes the signal which is dominated by stem (blue wavelength band). These voltage values are then sampled after a specified duration (225 us) and used for the dose calibration procedure.

Figure 4 c) and d) show the output for the maximum and minimum fiber configuration, respectively. Considering a nominal dose per pulse of 1 Gy for the FLASH beam and the 100pF feedback capacitor in the gated integrator, the output voltage can be calculated to be approximately ~0.6 V for each integrated LINAC pulse. This agrees with the values seen in Figure 4 c) and 4 d). For determination of the CLR, the output per pulse values of Ch 1 and Ch 2 were accumulated. The CLR value for this particular calibration was found to be $1.16 \pm 0.07$. Similarly, the gain value was found to be $3.09 \pm 0.21$. The standard deviation quoted here is based on 5 repeated measurements. Figure 4 c) shows the 5 repeated measures of a 35 pulse dose delivery sequence. The ramp-up period, which includes the first 5-6 pulses, is clearly visible. On average, the first 5 pulses delivered ~ $0.040 \pm 0.002$ Gy/Pulse. Once the calibration was performed, the next step was to characterize the W1 under FLASH dose-rates, since for any dose-based feedback mechanism to work, a dosimeter which is linear at high dose-rates and does not suffer from saturation issues in needed.

## 3.3 Exradin W1 Characterization

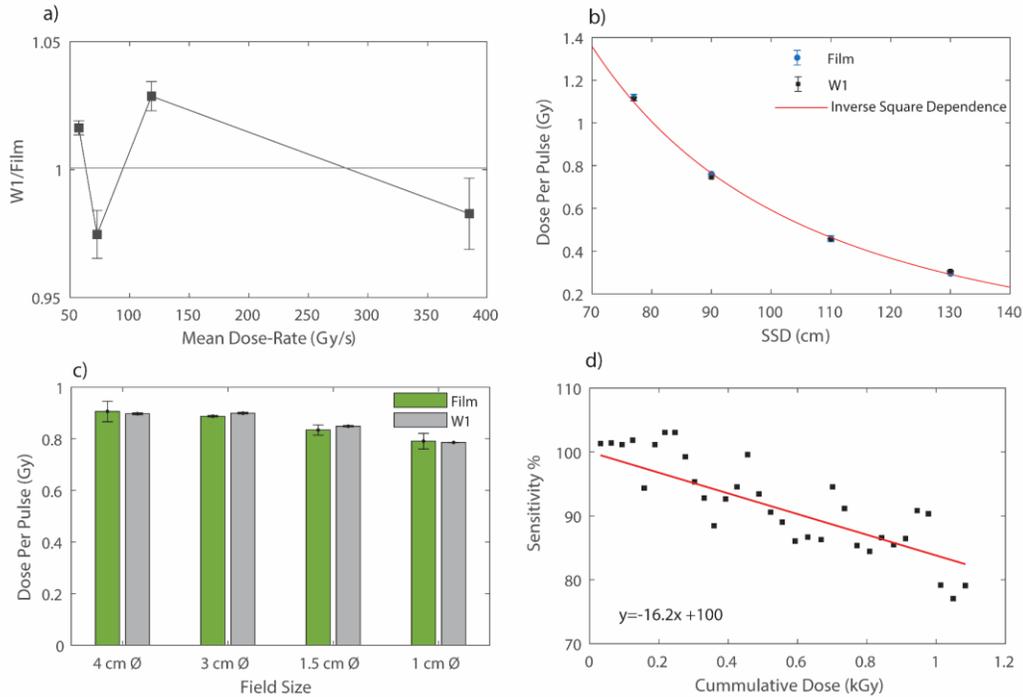

**Figure 5 a)** Mean dose-rate dependence of the W1 plastic scintillator. b) Dose-per pulse linearity of the W1. c) Field size characterization and d) shows the radiation damage characteristics of the W1.

The results of Exradin W1 characterization are presented in Figure 5. The W1 was found to be independent (within 3%) of any mean dose-rate effects when compared against film (Figure 5a). Since the Sync signal always pulses at 360 Hz, independent of the radiation pulses, a threshold value was used to only sample after a valid radiation pulse. The instantaneous dose-rate dependency, as shown in Figure 5b), yielded a close agreement between the two modalities, with an average difference of 2%, with film as the reference. The results for the field size dependency are shown in Figure 5 c). An average difference of 2.1%, with film as the reference, was seen between the two modalities indicating that the Cherenkov removal technique was accurately implemented. The radiation damage study, shown in Figure 5 d), indicates significant reduction in sensitivity of the dosimeter with increasing cumulative dose. A 16.2% loss in sensitivity per 1 kGy of dose was seen when the dosimeter was initially calibrated and characterized. However, with subsequent irradiations, a decrease in the rate of degradation was seen. For example, for the next 700 Gy, a decrease of 7% was observed. Therefore, for any experiments where accurate dosimetry is required, the W1 would have to be recalibrated just prior to the experiment, without any additional irradiations. Barring the radiation stability, the W1 was proven to be good candidate for online dose-based feedback. Therefore, the next step was to characterize the accuracy of the feedback system based on dose accumulation.

### 3.4 Dose-Based Feedback

**Table 1** Accuracy assessment of the dose-based feedback.

| Intended Dose (Gy) | 5 | 10 | 20 | 30 |
| --- | --- | --- | --- | --- |
| Controller Dose (Gy) | 5.2 ± 0.1 | 10.4 ± 0.2 | 20.5 ± 0.5 | 30.45 ± 0.2 |
| Film Dose (Gy) | 5.7 ± 0.1 | 11.1 ± 0.1 | 21.8 ± 0.7 | 31.7 ± 0.3 |
| Difference (Gy) | 0.48 ± 0.1 | 0.63 ± 0.2 | 1.25 ± 0.9 | 1.35 ± 0.4 |
| Num.Pulses | {15,14,14,16,15} | {28,24,28,29,25} | {42,38,38,43,42} | {56,53,53,54,58,58} |

For the dose-based feedback, the W1 was placed at iso-center in the calibration slab. To offset the radiation damage the W1 had accumulated prior to the study, a new calibration was obtained for this particular study. For the calibration, the built-in transmission chamber was used for beam control. Once calibrated, the internal feedback was turned off and the controller (via the dose-feedback implementation) was used as a beam gating signal. The accuracy of the feedback system was tested by delivering 5 Gy, 10 Gy, 20 Gy and 30 Gy with 5 repeated measures and verified against film data. The results of this study are shown in Table 1. It can be seen that controller provides accurate control over the dose delivery, if provided with the appropriate calibration. However, it can be seen in Table 1 that the final dose recorded by the film was on average ~ 0.9 Gy higher compared to the dose reading on the controller. Interestingly, the dose difference was lower for smaller dose deliveries. There are multiple reasons which contribute to the overdose reported by the film; 1) the calibration film was read-out immediately after exposure, but there is known time dependency of Gafchromic film, 2) the controller does not yet have the capability of modulating the pulse width of the last pulse in the dose sequence, which implies that accuracy of this method is as good as the nominal dose per pulse. For the 5 and 10 Gy delivery, the average dose delivered in the last pulse was 0.38 Gy and 0.68 Gy, respectively. For the 20 Gy and the 30 Gy delivery, the average dose per pulse in the last pulse were 0.75 Gy and 0.79 Gy, respectively. More importantly, it is worth noting that the number of pulses delivered were also recorded using the controller during this study. It can be seen that the number of pulses for similar doses vary significantly; a fact pointing towards the importance of dose based feedback instead of simple pulse counting circuitry.

Finally Figure 6, depicts the graphical user interface available to the user after a sample dose delivery. The interface displays information pertinent to the FLASH effect. In particular the output from the two channels and dose information is displayed in real-time. The number of delivered pulses is also displayed. The pulse count can also be inferred from the dose per pulse statistics, but the idea is to have redundant beam monitoring solutions. The pulse width of each radiation pulse in also displayed. The pulse width is measured in real-time for each pulse using the output from the RTU. For this particular acquisition, a wide 40 cm² field was used which suffers from minimal double pulsing issues and therefore reports accurate timing of the radiation pulses. Information about the pulse repetition rate is also displayed to the user. Note that there is a ramp-up for both the $D_p$ and the pulse width.

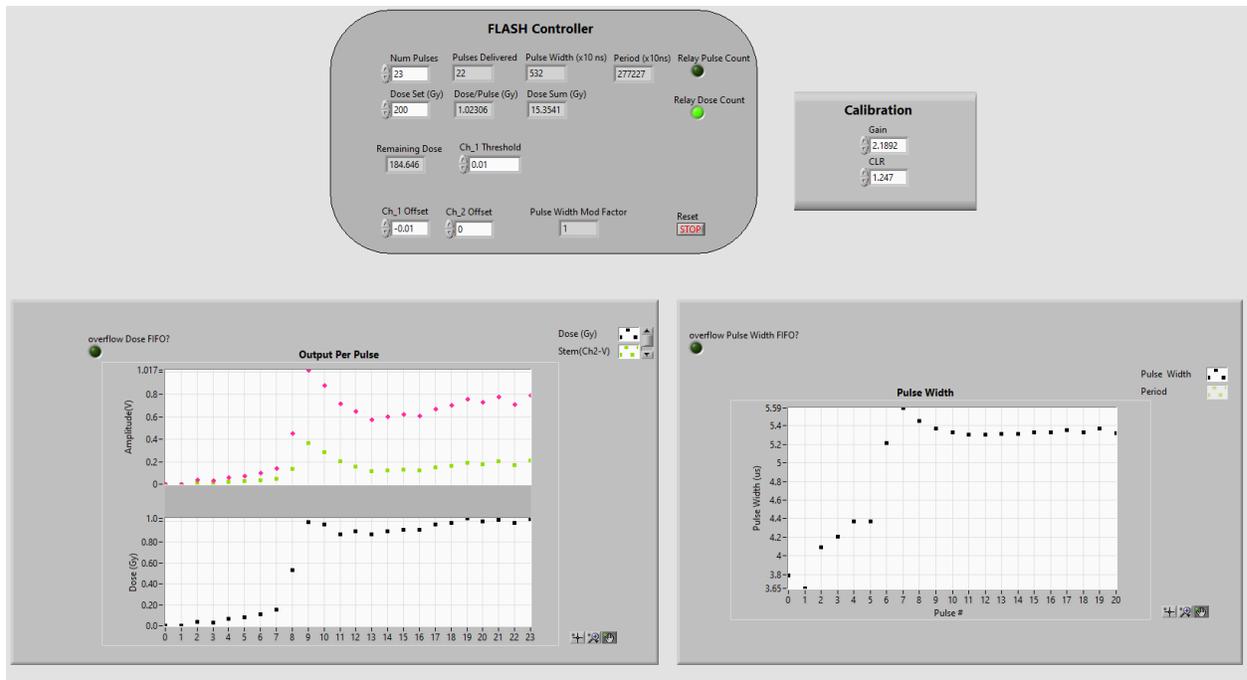

**Figure 6)** The user interface which is presented to the user after a sample dose irradiation. Pulse and dose counting metrics are presented from the gated integrator and in addition, information about the temporal structure of the beam is also presented to the user.

### 3.4 In-vivo Dosimetry

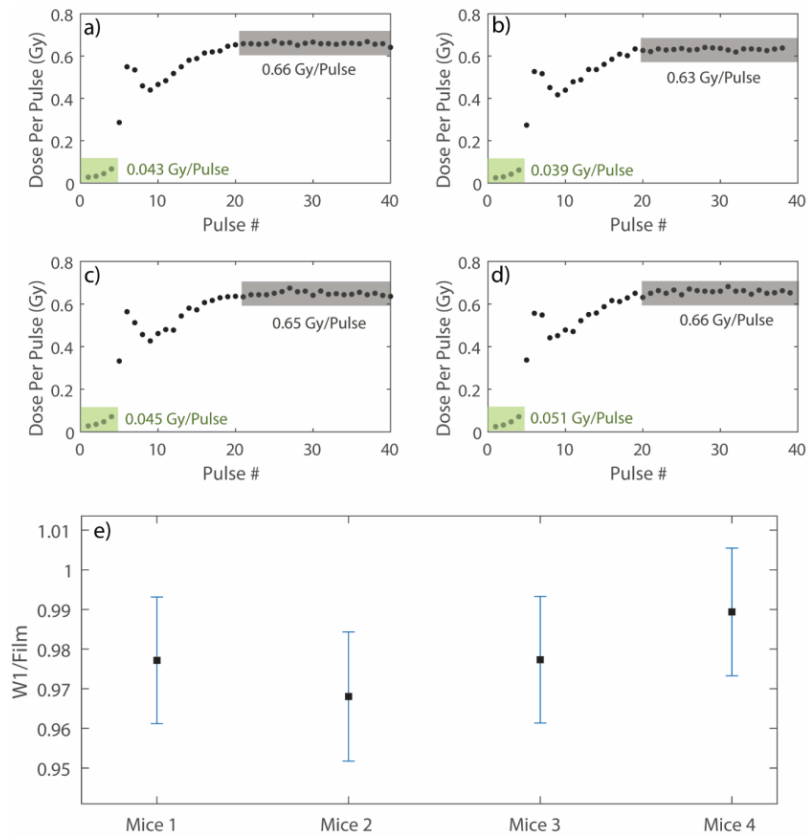

**Figure 7 a), b), c) and d)** In-vivo pulse-resolved dosimetry for the 4 mice. The green shaded areas signify the ramp-up period along with the average dose per pulse delivered in that domain. The grey shaded areas denote the stable output of the linac and the average dose per pulse delivered in that domain. **e)** Comparison of the cumulative dose against radiochromic film for the 4 mice.

As another example of the usefulness of the pulse-resolved beam monitoring system introduced in this study, in-vivo dosimetry results from 4 mice are presented in Figure 7. Note, only pulse counting feedback was used for this experiment. The ramp-up period was again clearly visible. With the 1 cm cut-out, the average dose delivered for the first 4 pulses across the 4 mice was again measured to be ~ 0.045 ± 0.004 Gy/Pulse. The dose/pulse eventually stabilized to a nominal value of around 0.65 ±0.01 Gy. When comparing the cumulative dose with radiochromic film, the W1 exhibited an under-response of ~3%. Since, it was seen earlier that the sensitivity of the detector deteriorates with dose, the dosimeter was calibrated prior to the in-vivo measurements with no additional doses delivered between the calibration setup and the actual experiment.

4. Discussions:

In this study, an FPGA-based hardware solution for real-time and pulse-resolved beam monitoring and feedback of FLASH-RT was introduced. A gated-integrator was used to accumulate dose in real-time. It should be noted that a gated integrator based solution has been previously proposed for pulse-resolved dosimetry[28–30], but none of the solutions employed an FPGA-based system and were not used in the context of beam monitoring and feedback for a FLASH enabled linac. Additionally, a commercially available plastic scintillator detector was coupled to the hardware and characterized at FLASH dose-rates. To the best of our knowledge, this is first instance of both dose-based feedback for FLASH-RT and characterization of the W1 at FLASH dose-rates. The W1 was found to be independent of any dose-rate effects (50- 350 Gy/s). The dose per pulse dependence, which is directly related to the instantaneous dose-rate, was also minimal at values ranging from 0.4-1.1 Gy/Pulse. The instantaneous dose-rate is considered an important parameter in eliciting the normal tissue-protective effect of FLASH. Therefore, the ability to quantify instantaneous dose-rate is of paramount importance. Additionally, no field size dependency was seen, indicating that the Cherenkov removal technique was sufficiently implemented.

Pulse counting and dose-based feedback were accurately implemented. This configuration is analogous to the two transmission chamber configuration typically used in a conventional linac as a form of redundant beam monitoring. The dose-accumulation based feedback was designed to take precedence, while the pulse-counting feedback was only used as a secondary form of feedback. The dose-based feedback was found to be accurate within 1 Gy, and more accurate results were seen at lower doses. For the next iteration, a calorimeter will be employed to calibrate the scintillator detector to absolute dose with high accuracy. Additionally, a solution is currently being worked on to modulate the linac pulse width, thereby giving an even better beam delivery system. It is expected that with these changes, the dose accuracy can be comparable to normal clinical tolerance.

Before luminescent dosimeters can be truly used as a source of real-time feedback to the LINAC, the issue of radiation degradation needs to be addressed. As seen in this study, radiation damage of a plastic scintillator can be a significant source of uncertainty. Essentially, each dosimeter fiber would need to be re-calibrated before irradiations to ensure accurate dose monitoring. Interestingly, the loss in sensitivity reported in this study was significantly higher compared to the values reported in literature for the Exradin W1. For example, an output loss of 1.6%/kGy for the first 5kGy, followed by a loss of 0.2%/ kGy for the next 20 kGy was reported by Dimitriadis et al[31]. Similarly, a reduction in output of approximately 0.28%/ kGy for the first 15 kGy, followed by a loss of 0.032%/ kGy for the next 100 kGy was reported by Carrasaco et al[32]. This is in stark contrast to the values reported in this study; an initial loss of 16%/ kGy. It was indeed seen that the rate of radiation damage did decrease with increasing cumulative dose (7% for the next 700 Gy). Radiation damage in plastic scintillators is a complex process mediated by oxygen diffusion, integrated dose and the dose-rate[33,34]. In general, the radiation damage of plastic scintillators can be modelled using an exponential of the form, $e^{\frac{-d}{D}}$ where d is the delivered dose, and D is the dose constant, which has been theorized to depend on dose-rate. It is generally agreed upon that it is not the damage to the fluorescent dopants, but rather the damage to the base material (polystyrene for the Exradin-W1), which is the dominant mechanism for reduction in output[35,36]. The base material can form absorption centers with increasing dose, which can lead to a loss in transmission[33,37]. The formation of absorption centers occur mainly via radical formation and the mobility of radicals is highly mediated by the presence of oxygen. For simplified experimental conditions, the formation of absorption centers in the base and its dependence on radical formation, oxygen diffusion length and dose-rate was quantified by Khachatryan et al[34] and it was reported that the dose constant, D is directly proportional to the square root of the dose-rate. The results presented in this study, contradict the conclusions of the aforementioned model, which would predict more radiation damage at low dose-rates due to increased oxygen diffusion. However, the authors only validated the model up to dose-rates of 0.05 Gy/s, far below the dose-rates commonly used in the FLASH regime. This discussion points towards the fact that there is a well-known dependence of dose-rate on radiation degradation of plastic scintillators. Nonetheless, the decrease in the rate of loss of sensitivity with increasing accumulated dose points toward the need to irradiate the dosimeter to a certain dose value, beyond which the decrease in sensitivity with increasing dose is relatively constant. To rule out the possibility of radiation damage to the gated integrator, which was also placed inside the radiation bunker, a constant voltage source coupled to a resistor was provided to the gated integrator during radiation delivery to test constancy of a constant current signal readout. No degradation of the integrator was seen after delivery of ~ 1000 pulses. Furthermore, within a single dose delivery, the presence of radiation did not interfere with the integrator. This is presented in Figure A1 in the supporting document where, the peak-peak voltage value of the integrated signal in the pre delivery, beam-on and post-delivery period was the same. Further exploration of radiation resistant scintillators would be ideal. In this regard, liquid based scintillators might be a viable solution. Moreover, it has been observed that for plastic scintillators, the change in transmission with increasing radiation damage occurs around 400-600 nm and remains relatively constant in the 600-800 nm range. Perhaps, red shifting dopants might also be an attractive choice. It should also be noted that the light guide can lead to radiation induced attenuation and fading. In general, high OH- content, pure silica light guides should be used, since they offer superior radiation resistance[38].

 Since the hardware is commercially available, and the FPGA was programmed using the LabView programming environment, the tools presented in this study can be implemented at other clinics and research groups conducting pre-clinical FLASH research. The code is available upon request in two formats; 1) the default LabView programming environment, and 2) as a VHDL code, which can be deployed to any FPGA board from Xilinx's Kintex-7 Family of FPGA boards. Importantly, the tools presented in this study are not limited to luminescence and can be used for other potentially attractive dosimeters for FLASH, such as micro-diamond detectors, making this hardware solution a flexible choice for online beam monitoring.

Finally, an under-response of ~3 % was seen for the in-vivo studies. The results are in agreement with previous findings using the W1 for in-vivo dose verification[39]. Alsanea et al[39] reported a ~4 % under-response at 37 °C relative to the calibration temperature. Although traditionally considered to be temperature independent, recent studies have found the plastic scintillators to be dependent on temperature[40]. Nonetheless, the ability to verify dose for each pulse is crucial to understanding in-vivo FLASH studies. For example, the first 4 pulses delivered on average 0.04 Gy/Pulse for the 1 cm cut-out used for the murine study and this type of information needs to be recorded in order to fully compare outcomes from different investigators. These values are comparable with conventional forms of radiotherapy and therefore, cannot be regarded as dose delivery under FLASH dose-rate regimes. Moreover, the quantities measured with this new control system will lead to more accurate reporting of FLASH effect and its relation to the temporal aspects of dose delivery.

5. **Conclusions:**

A real-time embedded industrial controller was presented for online beam monitoring and dose feedback control of an electron FLASH beam. Dose delivered within each pulse was measured and displayed in real-time. A commercially available scintillating detector was characterized under ultra-high dose-rate regime and was found to be suitable for dose verification, as long as routine recalibration are performed to account for the loss in sensitivity with radiation damage. Further iterations of the technique would include radiation resistant liquid scintillating detectors or calibration of the currently used versions for each irradiation session.

**Supporting Document**

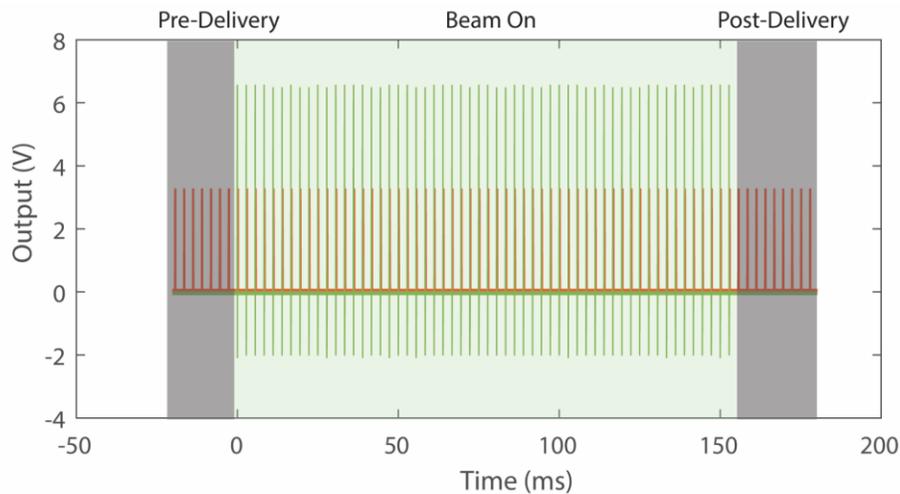

Figure A1 A constant voltage series coupled to a resistor was fed to the gated integrator. The output the integrator is shown in red and blue. To note, the blue output is hard to visualize, since it is overlaid on the red waveform. The green pulses are from the RTU, which denote the presence of a radiation pulse. The peak to peak value of the integrated voltage in the pre, post and beam on periods was found to be identical in all cases, indicating that the integrator is not affected by the presence of radiation and does not contribute to the loss in sensitivity. This test was repeated for a total of 20 repeated measurements and no loss in sensitivity and hence no amplifier damage was observed.

## Input/Output Modules

1) Digital I/O, NI-9401: The NI-9401 module is an 8-channel digital input or output 5V TTL module. The module is used to sense the Sync trigger from the linac and generates the reset, hold/integrates signals for the gated integrator.

2) ADC, NI-9252: The NI-9252 is an 8-channel analog input module which incorporates a separate 24-bit sigma delta ADC for each channel. The device offers a maximum sampling rate of 50,000 samples per second and +/- 10V measurement range. The module is connected to the two outputs of the gated integrator.

3) Counter, NI-9361: The NI-9361 module is an 8 channel 32 bit counter. The module is used for pulse counting but can be configured to measure pulse width and period of a pulse train as well.

The counter is connected to the RTU and measures the number of pulses (accounting for spurious pulses) and the pulse width for each pulse.

4) SSD Relay, NI-9485: The NI-9485 is an 8-channel solid state relay which is connected to the RPM Gating box.